\documentclass[aps,twocolumn]{revtex4}
\usepackage{bm}
\usepackage{graphicx}
\usepackage{epsfig}
\begin{document}

\title{Pure spin photocurrents in low-dimensional structures}
\author{S.A.~Tarasenko}\thanks{e-mail: tarasenko@coherent.ioffe.ru}
\author{E.L.~Ivchenko}
\affiliation{A.F.~Ioffe Physico-Technical Institute, Russian
Academy of Sciences, 194021 St.~Petersburg, Russia }
\begin{abstract}
As is well known the absorption of circularly polarized light in
semiconductors results in optical orientation of electron spins
and helicity-dependent electric photocurrent, and the absorption
of linearly polarized light is accompanied by optical alignment of
electron momenta. Here we show that the absorption of unpolarized
light leads to generation of a pure spin current, although both
the average electron spin and electric current vanish. We
demonstrate this for direct interband and intersubband as well as
indirect intraband (Drude-like) optical transitions in
semiconductor quantum wells (QWs).
\end{abstract}


\maketitle

\section{Introduction}
Spin and charge are among the basic properties of elementary
particles such as an electron, positron and proton. The
perturbation of a system of electrons by an electric field or
light may lead to a flow of the particles. The typical example is
an electric current that represents the directed flow of charge
carriers. Usually the electric currents do not entail a
considerable spin transfer because of the random orientation of
electron spins. However, the charge current can be accompanied by
a spin current if electron spins are co-oriented as it happens,
e.g., under injection of spin-polarized carriers from magnetic
materials~\cite{Fiederling,Ohno} or in the
optical-orientation-induced circular photogalvanic
effect~\cite{CPGE,CPGE2}. Furthermore, there exists a possibility
to create a pure spin current which is not accompanied by a net
charge transfer. This state represents a non-equilibrium
distribution when electrons with the spin ``up'' propagate mainly
in one direction and those with the spin ``down'' propagate in the
opposite direction. In terms of the kinetic theory, it can be
illustrated by a spin density matrix with two nonzero components,
$\rho_{s,s}(\bm{k})= \rho_{\bar{s},\bar{s}}(- \bm{k})$, where $s$
and $\bm{k}$ are the electron spin index and the wave vector, and
$\bar{s}$ means the spin opposite to $s$. Spin currents in
semiconductors can be driven by an electric field acting on
unpolarized free carriers which undergo a spin-dependent
scattering and/or propagate in a medium with spin-orbit coupling.
This is the so-called spin Hall effect where a pure spin transfer
appears in the direction perpendicular to the electric field,
see~\cite{Chazalviel,Bakun,Awschalom,Wunderlich} and references
therein. The spin currents can be induced as well by optical means
as a result of interference of one- and two-photon coherent
excitation with a two-color electromagnetic field~\cite{Hubner} or
under interband optical transitions in noncentrosymmetrical
semiconductors~~\cite{Bhat3D}.

Here we show that pure spin currents, accompanied neither by
charge transfer nor by spin orientation, can be achi\-ev\-ed under
absorption of linearly polarized or unpolarized light in
semiconductor low-dimensional systems. The effect is considered
here for direct interband and intersubband as well as for indirect
free-carrier optical transitions in semiconductor QWs.

Phenomenologically, the flux of electron spins is characterized by
a pseudotensor $\hat{\bm{F}}$ with the components
$F_{\beta}^{\alpha}$ describing the flow in the $\beta$ direction
of spins oriented along $\alpha$, with $\alpha$ and $\beta$ being
the Cartesian coordinates. In terms of the kinetic theory such a
component of the spin current is contributed by a non-equilibrium
correction $\propto \sigma_{\alpha} k_{\beta}$ to the electron
spin density matrix, where $\sigma_{\alpha}$ is the Pauli matrix.
In general, the concept of spin current is uncertain in systems
with spin-orbit interaction, since the spin and spin-dependent
velocity cannot be determined simultaneously~\cite{Rashba1}.
Mathematically it is caused by the fact that the Pauli matrices
and the velocity operator do not commute. However this problem is
overcome in the most of real semiconductor QWs where spin-orbit
interaction can be considered as a small perturbation. To the
first order in the constant of spin-orbit coupling and within the
relaxation time approximation, the pure spin current photoinduced
in the conduction band is given by
\begin{equation}\label{Spinflux}
F_{\alpha}^{\beta}=\sum_{\bm{k}} \tau_e \mathrm{Tr} \left[
\frac{\sigma_{\alpha}}{2} \, v_{\beta}(\bm{k}) \,\dot{\rho}
(\bm{k})\right]
\end{equation}
with the spin-dependent corrections being taken into account
either in the velocity operator $\bm{v}(\bm{k})$ or in the
photogeneration rate of the spin density matrix
$\dot{\rho}(\bm{k})$. Here $\tau_e$ is the relaxation time of the
spin current which can differ from the conventional momentum
relaxation time that governs the electron mobility.
Electron-electron collisions, which do not affect the mobility,
can contribute to $\tau_e$ as it happens, e.g., in the case of
spin relaxation~\cite{Glazov_ee}.

\section{Interband optical transitions}
Appearance of a pure spin current under direct optical transitions
is linked with two fundamental properties of semiconductor QWs,
namely, the linear in the wave vector $\bm k$ {\it spin splitting}
of energy spectrum and the spin-sensitive {\it selection rules}
for optical transitions~\cite{oo}. The effect is most easily
conceivable for direct transitions between the heavy-hole valence
subband $hh1$ and conduction subband $e1$ in QWs of the C$_s$
point symmetry, e.g. in (113)- or (110)-grown QWs. In such
structures the spin component along the QW normal $z$ is coupled
with the in-plane electron wave vector. This leads to
$\bm{k}$-linear spin-orbit splitting of the energy spectrum as
sketched in Fig.~1, where heavy hole subband $hh1$ is split into
two spin branches $\pm 3/2$. As a result they are shifted relative
to each other in the $\bm{k}$ space. In the reduced-symmetry
structures, the spin splitting of the conduction band  is usually
smaller than that of the valence band and not shown in Fig.~1 for
simplicity. Due to the selection rules the direct optical
transitions from the valence subband $hh1$ to the conduction
subband $e1$ can occur only if the electron angular momentum
changes by $\pm 1$. It follows then that the allowed transitions
are $|+3/2 \rangle \rightarrow |+1/2 \rangle$ and $|-3/2 \rangle
\rightarrow |-1/2 \rangle$, as illustrated in Fig.~1 by vertical
lines. Under excitation with linearly polarized or unpolarized
light the rates of both transitions are equal. In the presence of
spin splitting, the optical transitions induced by photons of the
fixed energy $\hbar\omega$ occur in the opposite points of the
$\bm{k}$ space for the spin branches $\pm 1/2$. The asymmetry of
photoexcitation results in a flow of electrons within each spin
branch. The corresponding fluxes $\bm{j}_{+ 1/2}$ and $\bm{j}_{-
1/2}$ are of equal strengths but of opposite directions. Thus,
this non-equilibrium electron distribution is characterized by the
nonzero spin current $\bm{j}_{\rm spin}$ = $(1/2) (\bm{j}_{+1/2} -
\bm{j}_{- 1/2})$ but a vanishing charge current, $e (\bm{j}_{+1/2}
+ \bm{j}_{- 1/2})=0$.

\begin{figure}[t]
\leavevmode \epsfxsize=0.75\linewidth
\centering{\epsfbox{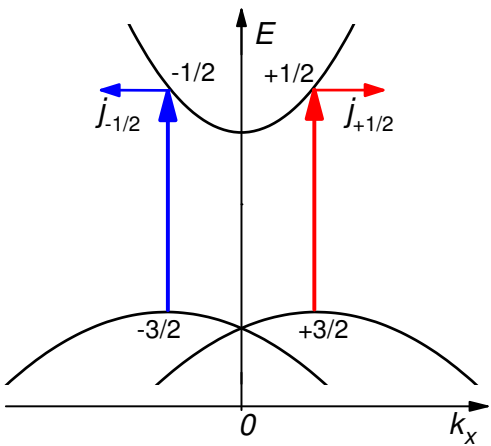}} \caption{Microscopic origin
of pure spin current induced by interband photoexcitation. The
vertical lines show the possible optical transitions.}
\end{figure}

The direction $\beta$ of the photoinduced spin current and the
orientation $\alpha$ of transmitted spins are determined by the
explicit form of spin-orbit interaction. The latter is governed by
the QW symmetry and can be varied. For QWs based on zinc blende
lattice semiconductors and grown along the crystallographic
direction $[110] \parallel z$, the light absorption leads to a
flow along $x
\parallel [1\bar{1}0]$ of spins oriented along $z$. This component
of the electron spin flow can be estimated as
\begin{equation}\label{j_inter}
F_{x}^{z} = \gamma_{zx}^{(hh1)} \frac{\tau_e}{2 \hbar}
\frac{m_h}{m_e+m_h} \frac{\eta_{cv}}{\hbar\omega} I \:,
\end{equation}
where $\gamma_{zx}^{(hh1)}$ is a constant describing the
$\bm{k}$-linear spin-orbit splitting of the $hh1$ subband, $m_e$
and $m_h$ are the electron and hole effective masses in the QW
plane, respectively, $\eta_{cv}$ is the light absorbance, and $I$
is the light intensity.

In (001)-grown QWs the absorption of linearly- or unpolarized
light results in a flow of electron spins oriented in the QW
plane. In contrast to the low-symmetry QWs considered above, in
(001)-QWs the $\bm{k}$-linear spin splitting of the $hh1$ valence
subband is suppressed and here, for the sake of simplicity, we
assume the parabolic spin-independent dispersion in the $hh1$
valence subband and take into account the spin-dependent
contribution
\begin{equation}\label{H_so}
H_{\mathrm{so}}^{(e1)} = \sum_{\alpha\beta} \gamma_{\alpha
\beta}^{(e1)} \, \sigma_{\alpha} k_{\beta}
\end{equation}
to the electron effective Hamiltonian. Then, to the first order in
the spin-orbit coupling, the components of the pure spin current
generated in the subband $e1$ are derived to be
\begin{equation} \label{intband}
F^{\alpha}_{\beta} = \gamma_{\alpha\beta}^{(e1)}
\frac{\tau_e}{2\hbar} \frac{m_e}{m_e+m_h} \frac{\eta_{cv}}{\hbar
\omega} I \:.
\end{equation}

\paragraph*{\textbf{Hole spin currents.}} For heavy holes $hh1$ one can introduce
the pseudospin description with pseudospins $\tilde{s} = \pm 1/2$
representing the hole states with the angular momentum $\pm 3/2$.
In addition to the electron spin current (\ref{intband}), a hole
pseudospin current is induced by interband photoexcitation with
linearly polarized light. In the geometry of normal incidence the
components $F^x_{\alpha}(hh1)$ and $F^y_{\alpha}(hh1)$ are given,
respectively, by the real and imaginary parts of
\[
F^x_{\alpha} + {\rm i} F^y_{\alpha} = (\gamma^{(e1)}_{x \alpha} +
{\rm i} \gamma^{(e1)}_{y \alpha}) \frac{\tau_h}{2\hbar}
\frac{m_e}{m_e + m_h} \frac{\eta_{cv}}{\hbar \omega} I (e_x + {\rm
i} e_y)^2 \:,
\]
where $\tau_h$ is the relaxation time of the hole spin current and
${\bm e}$ is the light polarization unit vector.

\section{Intersubband optical transitions}
The intersubband light absorption in $n$-doped QW structures is a
resonant process and occurs if the photon energy equals the energy
spacing between the subbands. In a simple one-band model, direct
optical transitions from the subband $e1$ to the subband $e2$ can
be induced only by irradiation with nonzero normal component $e_z$
of the polarization vector. These transitions occur with the spin
conservation, $(e1,+1/2)\rightarrow(e2,+1/2)$ and
$(e1,-1/2)\rightarrow(e2,-1/2)$, as illustrated in Fig.~2 by
vertical solid lines. Due to $\bm k$-linear spin splitting of the
$e1$ and $e2$ subbands, the optical transitions induced by photons
of the certain energy $\hbar\omega$ occur only at a fixed $k_x$
where the photon energy matches the energy separation between the
subbands~\cite{CPGE2}. Similarly to the interband excitation
considered in the previous section, these $k_x$-points are of
opposite sings for the spin branches $\pm 1/2$. Such
spin-dependent asymmetry of photoexcitation gives rise to pure
spin currents in both $e1$ and $e2$ subbands.

\begin{figure}[t]
\leavevmode \epsfxsize=0.95\linewidth
\centering{\epsfbox{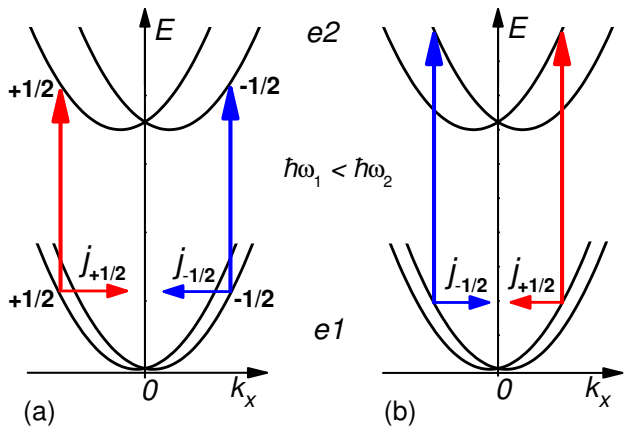}} \caption{Microscopic origin
of pure spin current induced by intersubband photoexcitation.
Change of the light frequency (Figs.~a,b) leads to reversal of the
spin current direction.}
\end{figure}

An interesting feature of the pure spin photocurrent induced under
intersubband transitions is its spectral behavior: an increase in
the photon energy $\hbar\omega$ (see Figs.~2a and 2b) leads to a
shift of the points $k_x$ that results in reversal of the spin
current direction. The explicit spectral dependence of the spin
photocurrent in ideal QWs depends on the specific fine structure
of the energy spectrum. However, in real structures the spectral
width of the intersubband resonance is broadened and hence
substantially exceeds the spectral width of the absorption
spectrum of an ideal structure. The broadening can be taken into
account assuming, e.g., that the energy separation $E_{21}$
between the subbands varies in the QW plane. Then to the first
order in the spin-orbit coupling the pure spin current generated
under intersubband optical transitions has the form
\begin{equation}\label{sf_intersubband}
F^{\alpha}_{\beta} = \frac{1}{2\hbar}
\left(\gamma^{(e2)}_{\alpha\beta}-\gamma^{(e1)}_{\alpha\beta}\right)
\frac{I}{\hbar\omega}
\end{equation}
\[
\times \left[ \tau_{e2}\, \eta_{21}(\hbar\omega) +
(\tau_{e1}-\tau_{e2}) \,\bar{E}\,
\frac{\,d\,\eta_{21}(\hbar\omega)}{d\,\hbar\omega} \right]  \:,
\]
where $\gamma^{(e1)}_{\alpha\beta}$ and
$\gamma^{(e2)}_{\alpha\beta}$ are the constants of the spin-orbit
coupling in the $e1$ and $e2$ subbands, $\tau_{e1}$ and
$\tau_{e2}$ are the corresponding relaxation times of the spin
currents, $\eta_{21}(\hbar\omega)$ is the intersubband light
absorbance with the inhomogeneous broadening being taken into
account, and $\bar{E}$ is the mean value of the electron kinetic
energy. It equals to $E_F/2$ for a 2D degenerate gas with the
Fermi energy $E_F$ and $k_B T$ for a 2D non-degenerate gas at the
temperature $T$. Intersubband light absorption is dominated by
spin-conserving optical transitions, therefore the pure spin
current~(\ref{sf_intersubband}) is proportional to the difference
of subband spin splittings. The spectral behavior of the spin
current is determined mainly by the second term in
Eq.~(\ref{sf_intersubband}) and repeats the derivative of the
light absorption spectrum $d\eta_{21}(\hbar\omega) / \hbar\omega$,
since the relaxation time in the excited subband $\tau_{e2}$ is
usually shorter than that in the lowest subband, $\tau_{e1}$.

\section{Free-carrier absorption}
Light absorption by free carriers, or the Drude-like absorption,
occurs in doped semiconductor structures when the photon energy
$\hbar\omega$ is smaller than the band gap as well as the
intersubband spacing. Because of the energy and momentum
conservation the free-carrier optical transitions become possible
if they are accompanied by electron scattering by acoustic or
optical phonons, static defects etc. Scattering-assisted
photoexcitation with linearly- or unpolarized light also gives
rise to a pure spin current. However, in contrast to the direct
transitions considered above, the spin splitting of the energy
spectrum leads to no essential contribution to the spin current
induced by free-carrier absorption. The more important
contribution comes from asymmetry of the electron spin-conserving
scattering. In semiconductor QWs the matrix element $V$ of
electron scattering by static defects or phonons has, in addition
to the main contribution $V_0$, an asymmetric spin-dependent
term~~\cite{Belinicher,bulli,AGW,IT_jetp}
\begin{equation}\label{V_asym}
V = V_0 + \sum_{\alpha\beta}V_{\alpha\beta} \,\sigma_{\alpha}
(k_{\beta} + k'_{\beta}) \:,
\end{equation}
where $\bm{k}$ and $\bm{k}'$ are the electron initial and
scattered wave vectors, respectively. Microscopically this
contribution is caused by the structural and bulk inversion
asymmetry similar to the Rashba/Dresselhaus spin splitting of the
electron subbands. The asymmetry of the electron-phonon
interaction results in nonequal rates of indirect optical
transitions for opposite wave vectors in each spin branch. This is
illustrated in Fig.~3, where the free-carrier absorption is shown
as a combined two-stage process involving electron-photon
interaction (vertical solid lines) and electron scattering
(dash\-ed horizontal lines). The scattering asymmetry is shown by
thick and thin dashed lines: electrons with the spin $+1/2$ are
preferably scattered into the states with $k_x
> 0$, while particles with the spin $-1/2$ are scattered
predominantly into the states with $k_x < 0$. The asymmetry causes
an imbalance in the distribution of photoexcited carriers in each
branch $s=\pm 1/2$ over the positive and negative $k_x$ states and
yields oppositely directed electron flows $\bm{j}_{\pm1/2}$ shown
by horizontal arrows. Similarly to the interband excitation
considered in the previous section, this non-equilibrium
distribution is characterized by a pure spin current without
charge transfer.

\begin{figure}[t]
\leavevmode \epsfxsize=0.75\linewidth
\centering{\epsfbox{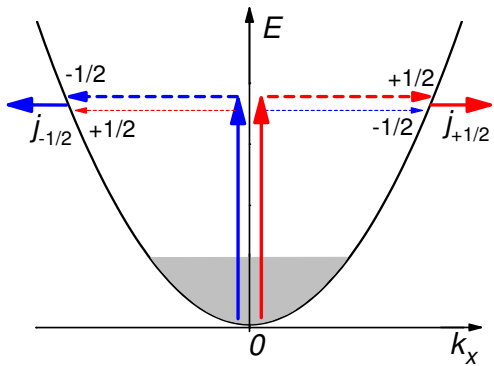}} \caption{Microscopic origin
of pure spin current induced under light absorption by free
electrons. The free-carrier absorption is a combined process
involving electron-photon interaction (vertical solid lines) and
electron scattering (dashed horizontal lines).}
\end{figure}

If the photon energy $\hbar\omega$ exceeds the typical electron
kinetic energy $\bar{E}$, the pure spin current induced by
free-carrier light absorption is given by
\begin{equation} \label{intraband}
F_{x}^{\alpha}=\frac{\tau_e}{\hbar} \left[ \frac{V_{\alpha
x}}{V_0} \left( 1+ \frac{|e_x|^2-|e_y|^2}{2} \right) +
\frac{V_{\alpha y}}{V_0}\ e_x e_y \right] \eta_{e1} I\ ,
\end{equation}
where $\eta_{e1}$ is the light absorbance in this spectral range.
The components $F_{y}^{\alpha}$ are obtained from
Eq.~(\ref{intraband}) by replacement $x \leftrightarrow y$.

In addition to the free-carrier absorption, the spin-de\-pen\-dent
asymmetry of electron-pho\-non interaction can also give rise to a
pure spin current in the process of energy relaxation of the
photoexcited electrons. In this relaxational mechanism the spin
current is generated in a system of hot carriers, independently of
heating means.

Mechanisms of pure spin photocurrent considered above can reveal
itself in an appearance of electric current in the presence of an
in-plane magnetic field. Indeed, the application of a magnetic
field results, due to the Zeeman effect, in different equilibrium
populations of the subbands. The currents ${\bm j}_{\pm 1/2}$
flowing in the opposite directions become non-equivalent which
results in a spin polarized net electric current~\cite{Ganichev}.

\paragraph*{\textbf{Valley-orbit current.}} In addition to the spin, free charge
carriers can be characterized by another internal property, e.g.,
by a well number in multi-QW structures or a valley index, $l$, in
multi-valley semiconductors. Thus, one can consider not only pure
spin currents but also pure orbit-valley currents in which case
the net electric current ${\bm j} = \sum_l {\bm j}_l$ vanishes but
the partial currents ${\bm j}_l$ contributed by carriers in the
$l$th valley are nonzero. In carbon nanotubes the index $l$ runs
through two equivalent one-dimensional subbands $(n, K)$ and $(-
n, K')$ formed near the $K$ and $K'$ valleys of the graphene sheet
rolled-up into a cylinder, where $n$ is the component of orbital
angular momentum along the tube principal axis $z$~\cite{Spivak}.
In chiral nanotubes the photoexcitation results in nonzero partial
flows, $j_z(K)$ and $j_z(K')$, which have opposite signs for
linearly polarized light. Another example is a GeSi/Si (111)-grown
quantum well structure. It has the overall C$_{3v}$ symmetry and
contains three equivalent two-dimensional valleys $l=1,2,3$. The
symmetry representing an individual valley is reduced to C$_s$ and
allows generation of a partial in-plane photocurrent ${\bm j}_l$
under normal light incidence. The net electric current is absent
but one can introduce the pure valley-orbit flows ${\bm j} =
\sum_l c_l {\bm j}_l$, where $c_l$ are arbitrary nonequal
coefficients.

\paragraph*{\textbf{Acknowledgement.}} We acknowledge helpful discussions
with V.V.~Bel'kov and S.D.~Ganichev. This work was supported by
the RFBR, the INTAS, programs of the RAS, and Foundation
``Dynasty'' - ICFPM.


\begin{thebibliography}{99}

\bibitem{Fiederling} R.~Fiederling, M.~Keim, G.~Reuscher, W.~Ossau, G.~Schmidt,
A.~Waag, and L.W.~Molenkamp, Nature {\bf 402}, 787 (1999).
\bibitem{Ohno} Y.~Ohno, D.K.~Young, B.~Beschoten, F.~Matsukura, H.~Ohno, and
D.D.~Awschalom, Nature {\bf 402}, 790 (1999).

\bibitem{CPGE} S.D.~Ganichev  and W.~Prettl,
J. Phys.: Condens. Matter {\bf 15}, R935 (2003).
\bibitem{CPGE2} S.D.~Ganichev, V.V.~Bel'kov, Petra~Schneider, E.L.~Ivchenko,
S.A.~Tarasenko, W.~Wegscheider, D.~Weiss, D.~Schuh,
E.V.~Beregulin, and W.~Prettl, Phys. Rev. B {\bf 68}, 035319
(2003).

\bibitem{Chazalviel} J.N.~Chazalviel and I.~Solomon, Phys.
Rev. Lett. {\bf 29}, 1676 (1972).
\bibitem{Bakun} A.A.~Bakun, B.P.~Zakharchenya, A.A.~Rogachev,
M.N.~Tkachuk, and V.G.~Fleisher, Pis'ma Zh. Eksp. Teor. Fiz. {\bf
40}, 464 (1984) [JETP Lett. {\bf 40}, 1293 (1984)].
\bibitem{Awschalom} Y.K.~Kato, R.C.~Myers, A.C.~Gossard, and
D.D.~Awschalom, Science {\bf 306}, 1910 (2004).
\bibitem{Wunderlich} J.~Wunderlich, B.~Kaestner, J.~Sinova, and
T.~Jungwirth, ArXiv:cond-mat/0410295.

\bibitem{Hubner} J.~H\"{u}bner, W.W.~R\"{u}hle, M.~Klude, D.~Hommel,
R.D.R.~Bhat, J.E.~Sipe, and H.M.~van~Driel, Phys. Rev. Lett. {\bf
90}, 216601 (2003).

\bibitem{Bhat3D} R.D.R.~Bhat, F.~Nastos, A.~Najmaie, and
J.E.~Sipe, arXiv:cond-mat/0404066.

\bibitem{Rashba1} E.I.~Rashba, Phys. Rev. B. {\bf 70}, 161201
(2004).

\bibitem{Glazov_ee} M.M.~Glazov and E.L.~Ivchenko, Pis'ma Zh.
Eksp. Teor. Fiz, {\bf 75}, 476 (2002) [JETP Lett. {\bf 75}, 403
(2002)].

\bibitem{oo} Optical Orientation, edited by F.~Meier and B.P.~Za\-khar\-chenya
(Elsevier Science, Amsterdam, 1984).

\bibitem{Belinicher} V.I.~Belinicher, Fiz. Tverd. Tela (Leningrad)
{\bf 24}, 15 (1982) [Sov. Phys. Solid
State {\bf 24}, 7 (1982)].
\bibitem{bulli} E.L.~Ivchenko and G.E. Pikus, Izv. Akad. Nauk
SSSR (ser. fiz.) {\bf 47}, 2369 (1983)
[Bull. Acad. Sci. USSR, Phys. Ser., {\bf 47}, 81 (1983)].
\bibitem{AGW} N.S.~Averkiev, L.E.~Golub, and M.~Willander, J.
Phys.: Condens. Matter {\bf 14}, R271 (2002).
\bibitem{IT_jetp} E.L.~Ivchenko and S.A.~Tarasenko, Zh. Eksp. Teor.
Fiz. {\bf 126}, 426 (2004) [JETP {\bf 99}, 379 (2004)].

\bibitem{Ganichev} V.V.~Bel'kov, S.D.~Ganichev, E.L.~Ivchenko,
S.A.~Tarasenko,  W.~Weber, S.~Giglberger,  M.~Olteanu,
H.-P.~Tranitz, S.N.~Danilov, Petra~Schneider, W.~Wegscheider,
D.~Weiss, and W.~Prettl, J. Phys.: Condens. Matter, to be
pulished.


\bibitem{Spivak} E.L.~Ivchenko and B.~Spivak, Phys. Rev. B {\bf 66},
155404 (2002).
\end{thebibliography}
\end{document}